\documentclass[aps,prl,preprint,showpacs,superscriptaddress]{revtex4-1}

\usepackage{graphicx}
\usepackage{amsmath}
\usepackage{bm}

\newcommand{\cmo}{CaMn$_7$O$_{12}$}
\newcommand{\tn}{\textit{T}$_{\mathrm{N}1}$}
\newcommand{\tnn}{\textit{T}$_{\mathrm{N}2}$}
\newcommand{\pvec}{\textit{\textbf{P}}}

\begin{document}

\title{Giant improper ferroelectricity in the ferroaxial magnet CaMn$_7$O$_{12}$}

\author{R. D. Johnson}
\email{r.johnson1@physics.ox.ac.uk}
\affiliation{Clarendon Laboratory, Department of Physics, University of Oxford, Oxford, OX1 3PU, United Kingdom}
\affiliation{ISIS facility, Rutherford Appleton Laboratory-STFC, Chilton, Didcot, OX11 0QX, United Kingdom}
\author{L. C. Chapon}
\affiliation{ISIS facility, Rutherford Appleton Laboratory-STFC, Chilton, Didcot, OX11 0QX, United Kingdom}
\affiliation{Institut Laue-Langevin, BP 156X, 38042 Grenoble, France}
\author{D. D. Khalyavin}
\author{P. Manuel}
\affiliation{ISIS facility, Rutherford Appleton Laboratory-STFC, Chilton, Didcot, OX11 0QX, United Kingdom}
\author{P. G. Radaelli}
\affiliation{Clarendon Laboratory, Department of Physics, University of Oxford, Oxford, OX1 3PU, United Kingdom}
\author{C. Martin}
\affiliation{Laboratoire CRISMAT, ENSICAEN, UMR F-6508 CNRS, 6 Boulevard du Marechal Juin, F-14050 Caen Cedex, France}

\date{\today}

\begin{abstract}

In rhombohedral \cmo, an improper ferroelectric polarization of magnitude 2870~$\mu$C~m$^{-2}$ is induced by an incommensurate helical magnetic structure that evolves below \tn~=~90~K. The electric polarization was found to be constrained to the high symmetry three-fold rotation axis of the crystal structure, perpendicular to the in-plane rotation of the magnetic moments. The multiferroicity is explained by the ferroaxial coupling mechanism, which in \cmo\ gives rise to the largest magnetically induced, electric polarization measured to date.
 
\end{abstract}


\maketitle

The coexistence of electric polarization and long-range magnetic order in single phase, multiferroic materials has been the topic of considerable research. The ultimate goal is to find a compound that displays a large improper ferroelectric polarization strongly coupled to the magnetic structure at high temperatures, and hence suitable for application in technology. A direct coupling between magnetism and ferroelectricity has been demonstrated in a number of multiferroics, for example TbMnO$_3$ \cite{kimura03}, TbMn$_2$O$_5$ \cite{hur04}, Ni$_3$V$_2$O$_8$ \cite{lawes05}, and MnWO$_4$ \cite{taniguchi06}, however the electric polarization tends to be small ($\sim10^2$~$\mu$C~m$^{-2}$) and the N$\acute{\mathrm{e}}$el temperature low ($<$~40~K). Few magneto-electric multiferroics, for example the hexaferrites \cite{kimura05} and CuO \cite{kimura08}, are polar close to room temperature (RT), and no such material has shown a large polarization comparable to that of proper ferroelectrics.

Here, we determine the origin of the magnetically induced ferroelectricity in \cmo\ \cite{zhang11} through the analysis of electric polarization, magnetic susceptibility, and neutron powder diffraction measurements. We show that the magnetically induced polarization, the largest  measured to date in any compound, develops at 90~K, oriented parallel to the \textit{c} axis. The improper ferroelectricity is induced by a helical magnetic structure with incommensurate propagation vector \textit{\textbf{k}}$_1$~=~(0,~1,~0.963). Contrary to conventional magnetic multiferroics, the electric polarization was found to lie perpendicular to the spin rotation plane, and is allowed through coupling to the ferroaxial component of the crystal structure \cite{johnson11}.

High quality single crystals of \cmo\ with approximate dimensions 300~x~300~x~300~$\mu$m were grown by the flux method. A 5~g mixture of CaCl$_2$:MnO$_2$ (weight ratio 1:3) was heated for 24~h in an alumina crucible in air at 850~$^\circ$C, and then cooled at a rate of 5$^\circ$~h$^{-1}$. X-ray diffraction and bulk property measurements showed that the single crystals were not twinned at RT. A 1~g powder sample was prepared by grinding single crystals and sieving through a 35~$\mu$m mesh. The electric polarization was determined by integrating the pyroelectric current measured on warming at a rate of 1~K~min$^{-1}$, having field cooled the sample in 4.4~kV~cm$^{-1}$. Magnetic susceptibility measurements were performed using a Quantum Design, Magnetic Properties Measurement System. Neutron powder diffraction (NPD) data were collected on the WISH time-of-flight diffractometer at ISIS, UK \cite{chapon11}. Data were collected at RT, and on warming between 1.5~K and 98~K in 3~K steps using a helium cryostat. The sample was enclosed in a cylindrical vanadium can. Magnetic structure refinements were performed using FullProf \cite{rodriguezcarvaja93} against data measured in detector banks at average $2\theta$ values of 58$^\circ$, 90$^\circ$, 122$^\circ$, and 154$^\circ$, each covering 32$^\circ$ of the scattering plane.

\begin{figure}
\includegraphics[width=8.5cm]{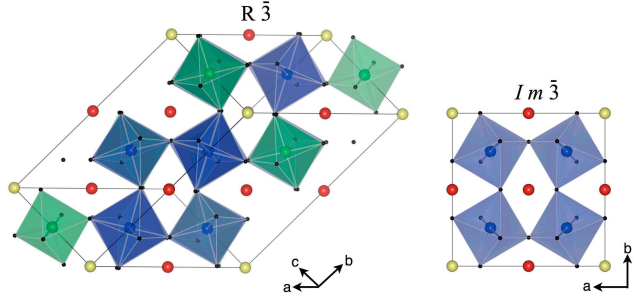}
\caption{\label{crystfig}(Color online) The crystal structure of \cmo\ in both hexagonal and cubic lattices. Calcium ions are shown in yellow, Mn1 in red, Mn2 in blue, Mn3 in green, and oxygen in black.}
\end{figure}

At high temperatures, \cmo\ crystallizes into a distorted perovskite structure with cubic space group $Im\bar{3}$ \cite{troyanchuk97} (see Fig. \ref{crystfig}). The crystal lattice undergoes a rhombohedral distortion on cooling through a first order phase transition at $\sim$440~K, giving the RT space group $R\bar{3}$ \cite{bochu80,przenioslo02} with lattice vectors ($-1$,~1,~0), (0,~$-1$,~1), and ($\frac{1}{2}$,~$\frac{1}{2}$,~$\frac{1}{2}$) with respect to those of the cubic unit cell. Here, all real and reciprocal space vectors are given in the hexagonal basis. In $R\bar{3}$ there are three symmetry inequivalent manganese sites that we label Mn1, Mn2, and Mn3. The Mn1 ions occupy sites of symmetry $\bar{1}$ (Wyckoff position $9e$) and form a 3:1 ordered occupation of the pseudo-cubic perovskite A-sites (the minority A-sites are occupied by non-magnetic calcium ions). Mn2 and Mn3 ions occupy sites of symmetry $\bar{1}$ and $\bar{3}$ (Wyckoff positions $9d$ and $3b$), respectively. They are octahedrally coordinated with oxygen and correspond to the pseudo-cubic B-sites with a ratio of three Mn2 to one Mn3. At RT Mn1 ions have a valence of +3, and the Mn2 and Mn3 ions are mixed-valent (+3.25). A further isostructural charge-ordering transition at 250~K  leaves nominal valences of +3 and +4 on the Mn2 and Mn3 ions, respectively \cite{przenioslo04}.

\begin{figure}
\includegraphics[width=8.5cm]{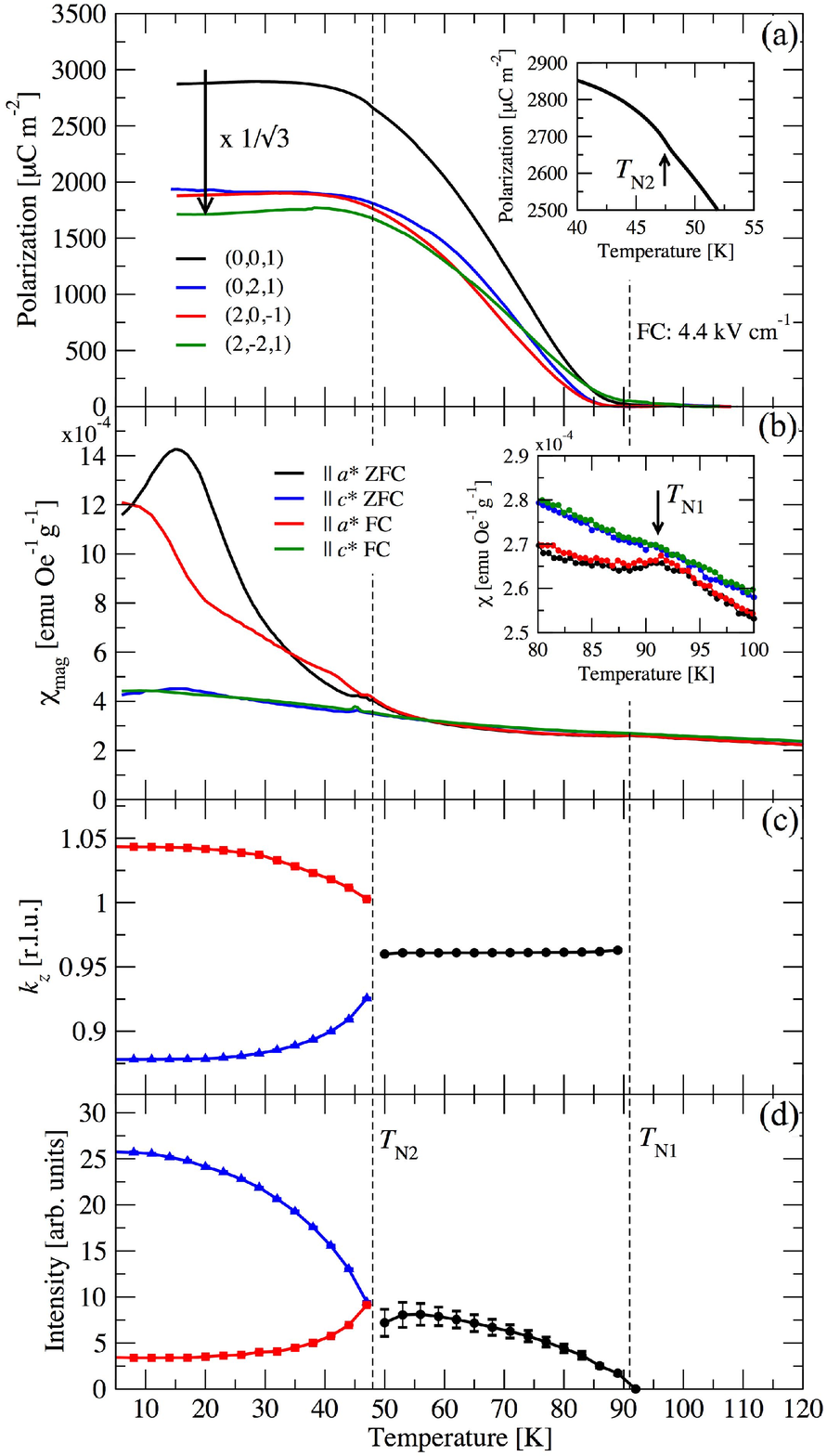}
\caption{\label{bulkfig}(Color online) a) The electric polarization of \cmo, measured parallel to the hexagonal \textit{c}-axis and the three pseudo-cubic $<$100$>$ axes. Inset: A small increase in \pvec\ at \tnn. b) The magnetic susceptibility parallel and perpendicular to the hexagonal \textit{c}-axis, measured with $H=500$~Oe, in both zero field cooled and field cooled (500~Oe) conditions. Inset: Reduction of the in-plane magnetic susceptibility at \tn. c) The variation of the incommensurate magnetic propagation vectors \textit{\textbf{k}}$_1$ (black circles), \textit{\textbf{k}}$_2$ (blue triangles), and \textit{\textbf{k}}$_3$ (red squares), refined against neutron powder diffraction data. The error bars are within the size of the data point. d) The integrated intensity of the (1,~-1,~1-$k_z$) magnetic reflection at $d\sim9\mathrm{\AA}$. The black circles, blue triangles and red squares correspond to \textit{\textbf{k}}$_1$, \textit{\textbf{k}}$_2$, and \textit{\textbf{k}}$_3$, respectively.}
\end{figure}

\cmo\ undergoes a further two phase transitions, \tn~=~90~K and \tnn~=~48~K, both of which have magnetic origin \cite{przenioslo99,sanchezandujar09}. Measurements of polycrystalline samples showed that the development of long-range antiferromagnetic order below \tn\ coincides with the onset of a large ferroelectric polarization \cite{zhang11}. In the present study, the electric polarization was measured on our single crystals perpendicular to the natural facets (parallel to the $<$100$>$ pseudo-cubic axes) and along the three-fold axis of the rhombohedral cell, as shown in Fig. \ref{bulkfig}a. The three measurements along (0,~2,~1), (2,~0,~-1), and (2,~-2,~1) showed polarization of equal magnitude within systematic error. In rhombohedral symmetry this equivalence may arise through one of two scenarios; either the electric polarization lies within the $ab$ plane, forming a 120$^\circ$ domain configuration, or the polarization is parallel to the three-fold $c$-axis. The fourth measurement along (0,~0,~1) conclusively demonstrates that the polarization vector, \pvec, lies parallel to the three-fold axis, where the polarization components along $<$0,~2,~1$>$ are reduced by a factor $1/\sqrt{3}$. Remarkably, the saturation value $|$\pvec$|$~=~2870~$\mu$C~m$^{-2}$ is the largest measured for magnetically-driven ferroelectrics to date, and is at least four times larger than that of the prototypal multiferroic, TbMnO$_3$ \cite{kimura03}.

The magnetic susceptibilities measured parallel and perpendicular to the \textit{c} axis (Fig. \ref{bulkfig}b) display accidents coinciding with  \tn\ and \tnn. At \tn, a small downturn of the in-plane susceptibility suggests antiferromagnetic ordering of the Mn moments in the hexagonal basal plane. At \tnn, the behavior is more complex with a small peak in the \textit{c} axis susceptibility, possibly indicative of an additional ordering along this direction, together with a rise of the in-plane susceptibility.

\begin{figure}
\includegraphics[width=8cm]{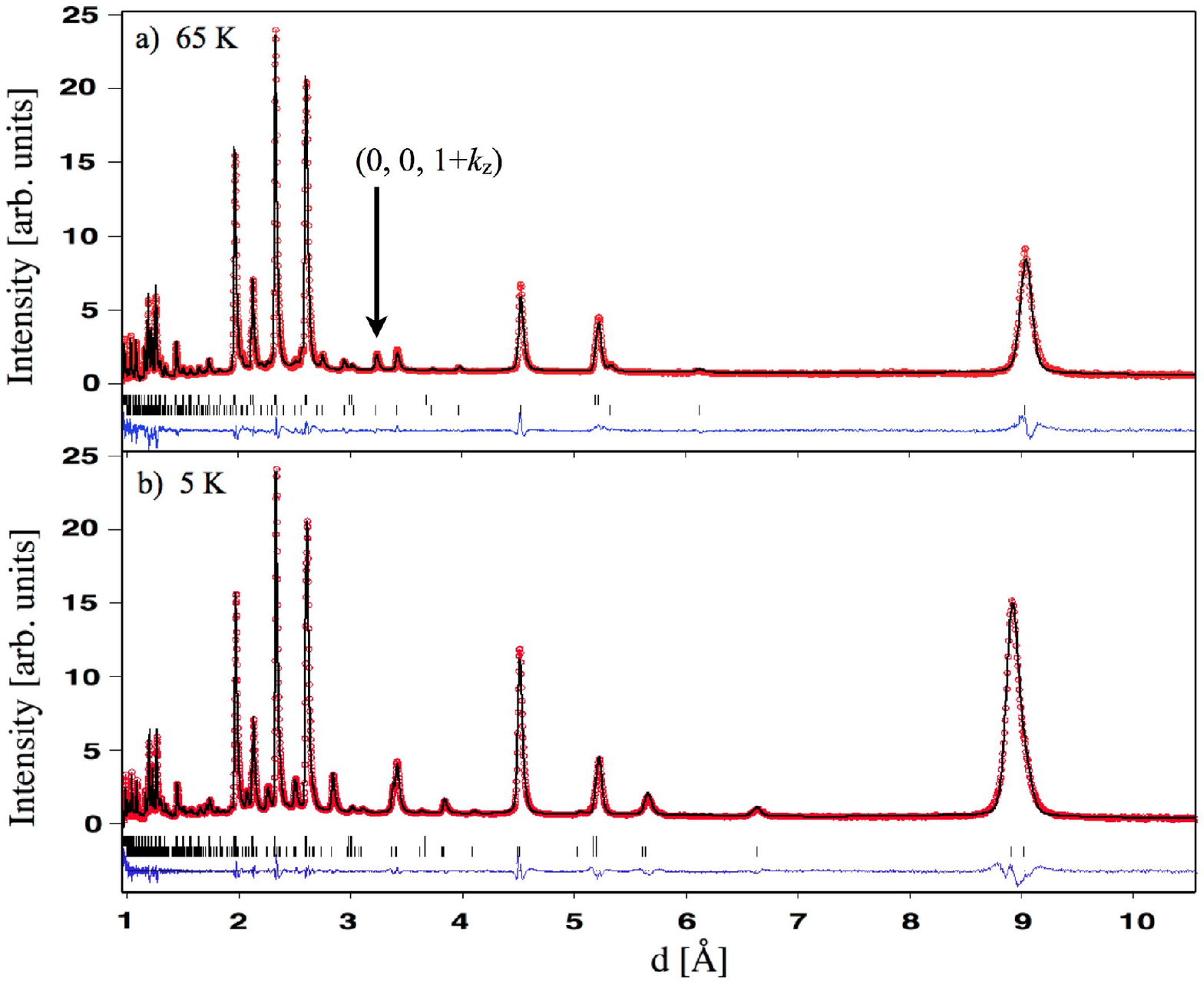}
\caption{\label{patternfig}(Color online) a) Rietveld refinement against neutron powder diffraction data at 65~K. b) Le Bail fit to data measured at 5~K. Top and bottom tick marks indicate the nuclear and magnetic peaks, respectively. The measured and calculated profiles are shown with dots and a continuous line, respectively. A difference curve (observed$-$calculated) is shown at the bottom.}
\end{figure}

Further insights into the nature of the magnetic state of \cmo\ were gained through NPD measurements. A large number ($>$15) of magnetic reflections were observed below \tn\ (Fig. \ref{patternfig}) indicative of long-range magnetic ordering. In the temperature range \tnn~$<T<$~\tn, all magnetic reflections can be indexed with the propagation vector \textit{\textbf{k}}$_1$~=~(0,~1,~0.963) (phase AFM~I). Below \tnn\ (AFM II) however, it is impossible to index all magnetic reflections with a single \textit{k}-vector. Instead, one observes the coexistence of two modulations along the same symmetry line of the Brillouin zone, with respective propagation vectors \textit{\textbf{k}}$_2$~=~(0,~1,~0.880) and \textit{\textbf{k}}$_3$~=~(0,~1,~1.042) at 5~K. In fact, all three propagation vectors lie along the same ($h_\mathrm{c}$,$h_\mathrm{c}$,$h_\mathrm{c}$) pseudo-cubic line of symmetry. The thermodiffractogram clearly shows that the magnetic peaks indexed by \textit{\textbf{k}}$_1$ above \tn\ split symmetrically at \tnn\ into the two contributions corresponding to \textit{\textbf{k}}$_2$=\textit{\textbf{k}}$_1$-(0,0,$\delta)$ and \textit{\textbf{k}}$_3$=\textit{\textbf{k}}$_1$+(0,0,$\delta$), as shown in Fig. \ref{bulkfig}c, as well as in the temperature dependence of the integrated intensities (Fig. \ref{bulkfig}d). We note that this magnetic phase diagram contradicts earlier neutron diffraction results which claimed a ferrimagnetic component in AFM I, incompatible with the bulk magnetization measurements \cite{przenioslo99}. 

\begin{figure}
\includegraphics[width=8.5cm]{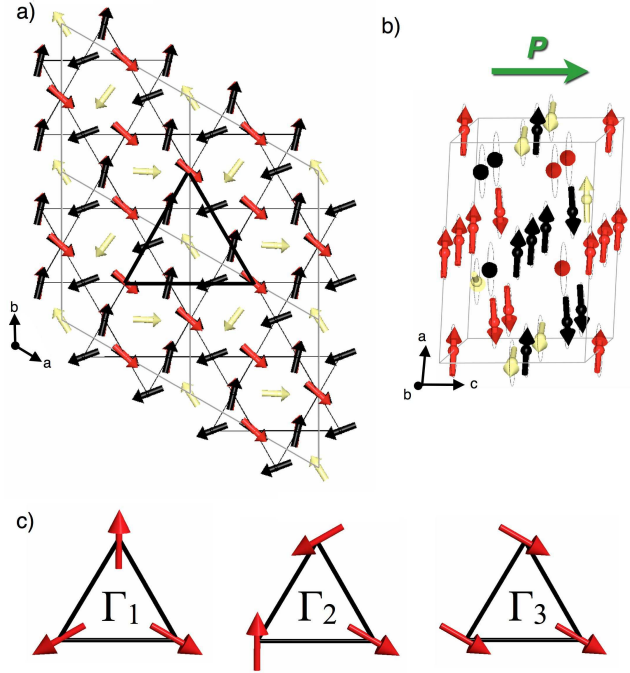}
\caption{\label{magfig}(Color online) (a), (b) The magnetic structure of \cmo\ in the \textit{ab} and \textit{ac} planes, respectively. Mn1, Mn2, and Mn3 ions are shown in red, black and yellow, respectively. The moments rotate in the \textit{ab}-plane with a circular envelope as depicted in (b), perpendicular to the electric polarization $\parallel$~\textit{c}-axis. c) The three triangular modes of symmetry equivalent Mn1 or Mn2 ions that correspond to $\Gamma_1$, $\Gamma_2$, and $\Gamma_3$. The magnetic structure of \cmo\ has $\Gamma_3$ symmetry, and the respective triangular mode of equivalent Mn1 sites is shown by the heavy line in (a).}
\end{figure}

In the following we focus on the magnetic structure of the AFM I phase, which is key to understanding the onset of the large electric polarization in \cmo. The magnetic structure of AFM II is more complex due to the presence of two \textit{k}-vectors and will require a single crystal, neutron study when larger specimens are available. For AFM I, symmetry analysis indicates that there are three irreducible representations (irreps) in the little group, all one-dimensional. All irreps appear in the decomposition of the magnetic representation for each site, suggesting that all Mn moments can be simultaneously ordered even if a single irrep is involved at the phase transition. For Mn1 and Mn2 sites, the possible magnetic arrangements spanned by the three irreps correspond to the symmetry-adapted modes of a system of spins on a triangle, as displayed in Fig. \ref{magfig}c. For a structure with equal moments, the $\Gamma_1$ and $\Gamma_2$ solutions lead to the extinction of the (0,~0,~l+\textit{k}$_z$) reflection due to the vanishing net moment in the triangle (null magnetic structure factor). However, such peaks are observed in the AFM I diffraction data (labelled in Fig. \ref{patternfig}a), dictating that the magnetic structure has $\Gamma_3$ symmetry, corresponding to a ferromagnetic arrangement of both Mn1 and Mn2 triangles, as shown in Fig. \ref{magfig}c. Refinement of the data at 65~K indicated that the moments are confined to lie in the \textit{ab} plane. The Mn$^{4+}$ moments follow a helical modulation along the \textit{c} axis with a circular envelop, imposed by symmetry. The Mn$^{3+}$ also follow a circular, helical modulation along $c$, since a refinement with an elliptical envelop (allowed by symmetry) did not lead to any improvement of the fit. The final refinement was conducted with five free parameters; the moment magnitudes of the three inequivalent manganese sites and the relative phases between each helix. The refinement is shown in Fig. \ref{patternfig}a and the results are given in Table \ref{magtab}. Reliability factors of $R_\mathrm{p}=6.65\%$ and $R_\mathrm{wp}=5.62\%$ were achieved for the 58$^\circ$ data, with a global (all banks) $\chi^2=12.0$.\\

The global magnetic structure can be described as follows: Mn$^{3+}$ ions form a buckled kagome lattice, as shown in Fig. \ref{magfig}a, in which the spin arrangement follows the so-called  ($\sqrt{3}$,$\sqrt{3}$) structure, one of the two possible ordered states of the classical kagome antiferromagnet \cite{harris92}. Each triangle has a 120$^{\circ}$ spin configuration. The buckled kagome layers are then stacked along the \textit{c} axis (Fig. \ref{magfig}b), with a long period modulation (\textit{k}$_z$=0.963) corresponding to a small global rotation of $\sim$ 4$^{\circ}$. Mn$^{4+}$ ions occupy the hexagonal voids of the kagome lattice and their moments are aligned ferromagnetically with respect to their nearest neighbor Mn$^{3+}$ ions in each layer. This chiral magnetic structure breaks inversion symmetry with magnetic point group $31'$.

The electric polarization was found to lie perpendicular to the spin rotation plane of the helical magnetic structure. This rules out the conventional inverse Dyzaloshinskii-Moriya \cite{sergienko06} or spin current \cite{katsura05} models for induced ferroelectricity, in which the electric polarization is constrained to the plane of rotation of the spins. In contrast, the magneto-electric properties of \cmo\ are allowed through coupling to the ferroaxial component of the crystal structure \cite{johnson11}, which also explains the multiferroic properties of other helical magnetic structures \cite{kimura06,arima07,kenzelmann07}. The nonpolar paramagnetic phase belongs to the $\bar{3}$ ferroaxial crystal class, for which a homogeneous structural rotation exists with respect to the high-temperature cubic group, represented by an axial vector \textit{\textbf{A}}. Symmetry considerations show that the chirality of the helical magnetic structure, $\sigma$~=~\textit{\textbf{r}}$_{ij}\cdot($\textit{\textbf{S}}$_i\times$\textit{\textbf{S}}$_j)$,  a time-even, parity-odd pseudoscalar, couples to \textit{\textbf{A}} and the polarization vector (necessarily $\parallel$  \textit{\textbf{A}}), \textit{\textbf{P$_z$}}, to form the invariant $\gamma$\textit{\textbf{P$_z$}}$\sigma$\textbf{A}, where $\gamma$ is a coupling constant. Considering the dielectric term in the free energy expansion, $\frac{1}{\epsilon_{zz}}$\textit{\textbf{P}}$_z^2$, the stability condition leads to \textit{\textbf{P$_z$}}$\propto$ -$\epsilon_{zz} \sigma$\textit{\textbf{A}}. In the first approximation, \textit{\textbf{A}} is constant at the temperatures in question (well below the critical temperature of the structural transition), therefore \textit{\textbf{P}}$_z$ varies linearly with $\sigma$. In the mean-field approximation the magnetic critical exponent is 0.5. $\sigma$, and hence \textit{\textbf{P}}$_z$, should vary as the square of the magnetization, (\textit{\textbf{S}}$_i\times$\textit{\textbf{S}}$_j$), and therefore linearly with temperature, as observed in the temperature dependence of the electrical polarization in \cmo\ (except very close to the transition).

\begin{table}
\caption{\label{magtab}Magnetic structure parameters at 65~K}
\begin{ruledtabular}
\begin{tabular}{c c c c c c c}
Atom & Valence & \textit{x} & \textit{y} & \textit{z} & $|$\textit{\textbf{M}}$|$ ($\mu_\mathrm{B}$) & $\phi$ ($\pi$ rad)\\
\hline
Mn1 & +3 & 0.5 & 0 & 0 & 2.31(3) & 0\\
Mn2 & +3 & 0.5 & 0 & 0.5 & 2.33(3) & -0.03(5)\\
Mn3 & +4 & 0 & 0 & 0.5 & 2.01(8) & 0.84(4)\\
\end{tabular}
\end{ruledtabular}
\end{table}

The exact microscopic coupling between spins and electric dipoles is unclear due to the complexity associated with the numerous competing exchange interactions. However, it is likely that exchange-striction coupled to the Mn$^{3+}$ / Mn$^{4+}$ charge order is responsible for such a large electric polarization.

To conclude, we have found that in \cmo\ a giant improper ferroelectric polarization is induced parallel to the \textit{c} axis by an in-plane helical structure at \tn~=~90~K. The multiferroicity can be explained by the ferroaxial coupling of the magnetic chirality to the macroscopic structural rotations associated with the $R\bar3$ paramagnetic phase.

\begin{acknowledgments}
The work done at the University of Oxford was funded by an EPSRC grant, number EP/J003557/1, entitled ``New Concepts in Multiferroics and Magnetoelectrics".
\end{acknowledgments}

\bibliography{cmo}

\begin{thebibliography}{22}%
\makeatletter
\providecommand \@ifxundefined [1]{%
 \@ifx{#1\undefined}
}%
\providecommand \@ifnum [1]{%
 \ifnum #1\expandafter \@firstoftwo
 \else \expandafter \@secondoftwo
 \fi
}%
\providecommand \@ifx [1]{%
 \ifx #1\expandafter \@firstoftwo
 \else \expandafter \@secondoftwo
 \fi
}%
\providecommand \natexlab [1]{#1}%
\providecommand \enquote  [1]{``#1''}%
\providecommand \bibnamefont  [1]{#1}%
\providecommand \bibfnamefont [1]{#1}%
\providecommand \citenamefont [1]{#1}%
\providecommand \href@noop [0]{\@secondoftwo}%
\providecommand \href [0]{\begingroup \@sanitize@url \@href}%
\providecommand \@href[1]{\@@startlink{#1}\@@href}%
\providecommand \@@href[1]{\endgroup#1\@@endlink}%
\providecommand \@sanitize@url [0]{\catcode `\\12\catcode `\$12\catcode
  `\&12\catcode `\#12\catcode `\^12\catcode `\_12\catcode `\%12\relax}%
\providecommand \@@startlink[1]{}%
\providecommand \@@endlink[0]{}%
\providecommand \url  [0]{\begingroup\@sanitize@url \@url }%
\providecommand \@url [1]{\endgroup\@href {#1}{\urlprefix }}%
\providecommand \urlprefix  [0]{URL }%
\providecommand \Eprint [0]{\href }%
\providecommand \doibase [0]{http://dx.doi.org/}%
\providecommand \selectlanguage [0]{\@gobble}%
\providecommand \bibinfo  [0]{\@secondoftwo}%
\providecommand \bibfield  [0]{\@secondoftwo}%
\providecommand \translation [1]{[#1]}%
\providecommand \BibitemOpen [0]{}%
\providecommand \bibitemStop [0]{}%
\providecommand \bibitemNoStop [0]{.\EOS\space}%
\providecommand \EOS [0]{\spacefactor3000\relax}%
\providecommand \BibitemShut  [1]{\csname bibitem#1\endcsname}%
\let\auto@bib@innerbib\@empty
\bibitem [{\citenamefont {Kimura}\ \emph {et~al.}(2003)\citenamefont {Kimura},
  \citenamefont {Goto}, \citenamefont {Shintani}, \citenamefont {Ishizaka},
  \citenamefont {Arima},\ and\ \citenamefont {Tokura}}]{kimura03}%
  \BibitemOpen
  \bibfield  {author} {\bibinfo {author} {\bibfnamefont {T.}~\bibnamefont
  {Kimura}}, \bibinfo {author} {\bibfnamefont {T.}~\bibnamefont {Goto}},
  \bibinfo {author} {\bibfnamefont {H.}~\bibnamefont {Shintani}}, \bibinfo
  {author} {\bibfnamefont {K.}~\bibnamefont {Ishizaka}}, \bibinfo {author}
  {\bibfnamefont {T.}~\bibnamefont {Arima}}, \ and\ \bibinfo {author}
  {\bibfnamefont {Y.}~\bibnamefont {Tokura}},\ }\href@noop {} {\bibfield
  {journal} {\bibinfo  {journal} {Nature}\ }\textbf {\bibinfo {volume} {426}},\
  \bibinfo {pages} {55} (\bibinfo {year} {2003})}\BibitemShut {NoStop}%
\bibitem [{\citenamefont {Hur}\ \emph {et~al.}(2004)\citenamefont {Hur},
  \citenamefont {Park}, \citenamefont {Sharma}, \citenamefont {Ahn},
  \citenamefont {Guha},\ and\ \citenamefont {Cheong}}]{hur04}%
  \BibitemOpen
  \bibfield  {author} {\bibinfo {author} {\bibfnamefont {N.}~\bibnamefont
  {Hur}}, \bibinfo {author} {\bibfnamefont {S.}~\bibnamefont {Park}}, \bibinfo
  {author} {\bibfnamefont {P.~A.}\ \bibnamefont {Sharma}}, \bibinfo {author}
  {\bibfnamefont {J.~S.}\ \bibnamefont {Ahn}}, \bibinfo {author} {\bibfnamefont
  {S.}~\bibnamefont {Guha}}, \ and\ \bibinfo {author} {\bibfnamefont {S.-W.}\
  \bibnamefont {Cheong}},\ }\href@noop {} {\bibfield  {journal} {\bibinfo
  {journal} {Nature}\ }\textbf {\bibinfo {volume} {429}},\ \bibinfo {pages}
  {392} (\bibinfo {year} {2004})}\BibitemShut {NoStop}%
\bibitem [{\citenamefont {Lawes}\ \emph {et~al.}(2005)\citenamefont {Lawes},
  \citenamefont {Harris}, \citenamefont {Kimura}, \citenamefont {Rogado},
  \citenamefont {Cava}, \citenamefont {Aharony}, \citenamefont {Entin-Wohlman},
  \citenamefont {Yildrim}, \citenamefont {Kenzelmann}, \citenamefont
  {Broholm},\ and\ \citenamefont {Ramirez}}]{lawes05}%
  \BibitemOpen
  \bibfield  {author} {\bibinfo {author} {\bibfnamefont {G.}~\bibnamefont
  {Lawes}}, \bibinfo {author} {\bibfnamefont {A.~B.}\ \bibnamefont {Harris}},
  \bibinfo {author} {\bibfnamefont {T.}~\bibnamefont {Kimura}}, \bibinfo
  {author} {\bibfnamefont {N.}~\bibnamefont {Rogado}}, \bibinfo {author}
  {\bibfnamefont {R.~J.}\ \bibnamefont {Cava}}, \bibinfo {author}
  {\bibfnamefont {A.}~\bibnamefont {Aharony}}, \bibinfo {author} {\bibfnamefont
  {O.}~\bibnamefont {Entin-Wohlman}}, \bibinfo {author} {\bibfnamefont
  {T.}~\bibnamefont {Yildrim}}, \bibinfo {author} {\bibfnamefont
  {M.}~\bibnamefont {Kenzelmann}}, \bibinfo {author} {\bibfnamefont
  {C.}~\bibnamefont {Broholm}}, \ and\ \bibinfo {author} {\bibfnamefont
  {A.~P.}\ \bibnamefont {Ramirez}},\ }\href@noop {} {\bibfield  {journal}
  {\bibinfo  {journal} {Phys. Rev. Lett.}\ }\textbf {\bibinfo {volume} {95}},\
  \bibinfo {pages} {087205} (\bibinfo {year} {2005})}\BibitemShut {NoStop}%
\bibitem [{\citenamefont {Taniguchi}\ \emph {et~al.}(2006)\citenamefont
  {Taniguchi}, \citenamefont {Abe}, \citenamefont {Takenobu}, \citenamefont
  {Iwasa},\ and\ \citenamefont {Arima}}]{taniguchi06}%
  \BibitemOpen
  \bibfield  {author} {\bibinfo {author} {\bibfnamefont {K.}~\bibnamefont
  {Taniguchi}}, \bibinfo {author} {\bibfnamefont {N.}~\bibnamefont {Abe}},
  \bibinfo {author} {\bibfnamefont {T.}~\bibnamefont {Takenobu}}, \bibinfo
  {author} {\bibfnamefont {Y.}~\bibnamefont {Iwasa}}, \ and\ \bibinfo {author}
  {\bibfnamefont {T.}~\bibnamefont {Arima}},\ }\href@noop {} {\bibfield
  {journal} {\bibinfo  {journal} {Phys. Rev. Lett.}\ }\textbf {\bibinfo
  {volume} {97}},\ \bibinfo {pages} {097203} (\bibinfo {year}
  {2006})}\BibitemShut {NoStop}%
\bibitem [{\citenamefont {Kimura}\ \emph {et~al.}(2005)\citenamefont {Kimura},
  \citenamefont {Lawes},\ and\ \citenamefont {Ramirez}}]{kimura05}%
  \BibitemOpen
  \bibfield  {author} {\bibinfo {author} {\bibfnamefont {T.}~\bibnamefont
  {Kimura}}, \bibinfo {author} {\bibfnamefont {G.}~\bibnamefont {Lawes}}, \
  and\ \bibinfo {author} {\bibfnamefont {A.~P.}\ \bibnamefont {Ramirez}},\
  }\href@noop {} {\bibfield  {journal} {\bibinfo  {journal} {Phys. Rev. Lett.}\
  }\textbf {\bibinfo {volume} {94}},\ \bibinfo {pages} {137201} (\bibinfo
  {year} {2005})}\BibitemShut {NoStop}%
\bibitem [{\citenamefont {Kimura}\ \emph {et~al.}(2008)\citenamefont {Kimura},
  \citenamefont {Sekio}, \citenamefont {Nakamura}, \citenamefont {Siegrist},\
  and\ \citenamefont {Ramirez}}]{kimura08}%
  \BibitemOpen
  \bibfield  {author} {\bibinfo {author} {\bibfnamefont {T.}~\bibnamefont
  {Kimura}}, \bibinfo {author} {\bibfnamefont {Y.}~\bibnamefont {Sekio}},
  \bibinfo {author} {\bibfnamefont {H.}~\bibnamefont {Nakamura}}, \bibinfo
  {author} {\bibfnamefont {T.}~\bibnamefont {Siegrist}}, \ and\ \bibinfo
  {author} {\bibfnamefont {A.~P.}\ \bibnamefont {Ramirez}},\ }\href@noop {}
  {\bibfield  {journal} {\bibinfo  {journal} {Nature Materials}\ }\textbf
  {\bibinfo {volume} {7}},\ \bibinfo {pages} {291} (\bibinfo {year}
  {2008})}\BibitemShut {NoStop}%
\bibitem [{\citenamefont {Zhang}\ \emph {et~al.}(2011)\citenamefont {Zhang},
  \citenamefont {Dong}, \citenamefont {Yan}, \citenamefont {Zhang},
  \citenamefont {Yunoki}, \citenamefont {Dagotto},\ and\ \citenamefont
  {Liu}}]{zhang11}%
  \BibitemOpen
  \bibfield  {author} {\bibinfo {author} {\bibfnamefont {G.}~\bibnamefont
  {Zhang}}, \bibinfo {author} {\bibfnamefont {S.}~\bibnamefont {Dong}},
  \bibinfo {author} {\bibfnamefont {Z.}~\bibnamefont {Yan}}, \bibinfo {author}
  {\bibfnamefont {Q.}~\bibnamefont {Zhang}}, \bibinfo {author} {\bibfnamefont
  {S.}~\bibnamefont {Yunoki}}, \bibinfo {author} {\bibfnamefont
  {E.}~\bibnamefont {Dagotto}}, \ and\ \bibinfo {author} {\bibfnamefont
  {J.-M.}\ \bibnamefont {Liu}},\ }\href@noop {} {\bibfield  {journal} {\bibinfo
   {journal} {arXiv:1101.2276v1}\ } (\bibinfo {year} {2011})}\BibitemShut
  {NoStop}%
\bibitem [{\citenamefont {Johnson}\ \emph {et~al.}(2011)\citenamefont
  {Johnson}, \citenamefont {Nair}, \citenamefont {Chapon}, \citenamefont
  {Bombardi}, \citenamefont {Vecchini}, \citenamefont {Prabhakaran},
  \citenamefont {Boothroyd},\ and\ \citenamefont {Radaelli}}]{johnson11}%
  \BibitemOpen
  \bibfield  {author} {\bibinfo {author} {\bibfnamefont {R.~D.}\ \bibnamefont
  {Johnson}}, \bibinfo {author} {\bibfnamefont {S.}~\bibnamefont {Nair}},
  \bibinfo {author} {\bibfnamefont {L.~C.}\ \bibnamefont {Chapon}}, \bibinfo
  {author} {\bibfnamefont {A.}~\bibnamefont {Bombardi}}, \bibinfo {author}
  {\bibfnamefont {C.}~\bibnamefont {Vecchini}}, \bibinfo {author}
  {\bibfnamefont {D.}~\bibnamefont {Prabhakaran}}, \bibinfo {author}
  {\bibfnamefont {A.~T.}\ \bibnamefont {Boothroyd}}, \ and\ \bibinfo {author}
  {\bibfnamefont {P.~G.}\ \bibnamefont {Radaelli}},\ }\href@noop {} {\bibfield
  {journal} {\bibinfo  {journal} {Phys. Rev. Lett.}\ }\textbf {\bibinfo
  {volume} {107}},\ \bibinfo {pages} {137205} (\bibinfo {year}
  {2011})}\BibitemShut {NoStop}%
\bibitem [{\citenamefont {Chapon}\ \emph {et~al.}(2011)\citenamefont {Chapon},
  \citenamefont {Manuel}, \citenamefont {Radaelli}, \citenamefont {Benson},
  \citenamefont {Perrott}, \citenamefont {Ansell}, \citenamefont {Rhodes},
  \citenamefont {Raspino}, \citenamefont {Duxbury}, \citenamefont {Spill},\
  and\ \citenamefont {Norris}}]{chapon11}%
  \BibitemOpen
  \bibfield  {author} {\bibinfo {author} {\bibfnamefont {L.~C.}\ \bibnamefont
  {Chapon}}, \bibinfo {author} {\bibfnamefont {P.}~\bibnamefont {Manuel}},
  \bibinfo {author} {\bibfnamefont {P.~G.}\ \bibnamefont {Radaelli}}, \bibinfo
  {author} {\bibfnamefont {C.}~\bibnamefont {Benson}}, \bibinfo {author}
  {\bibfnamefont {L.}~\bibnamefont {Perrott}}, \bibinfo {author} {\bibfnamefont
  {S.}~\bibnamefont {Ansell}}, \bibinfo {author} {\bibfnamefont {N.~J.}\
  \bibnamefont {Rhodes}}, \bibinfo {author} {\bibfnamefont {D.}~\bibnamefont
  {Raspino}}, \bibinfo {author} {\bibfnamefont {D.}~\bibnamefont {Duxbury}},
  \bibinfo {author} {\bibfnamefont {E.}~\bibnamefont {Spill}}, \ and\ \bibinfo
  {author} {\bibfnamefont {J.}~\bibnamefont {Norris}},\ }\href@noop {}
  {\bibfield  {journal} {\bibinfo  {journal} {Neutron News}\ }\textbf {\bibinfo
  {volume} {22}},\ \bibinfo {pages} {22} (\bibinfo {year} {2011})}\BibitemShut
  {NoStop}%
\bibitem [{\citenamefont {Rodr{\'\i}guez-Carvajal}(1993)}]{rodriguezcarvaja93}%
  \BibitemOpen
  \bibfield  {author} {\bibinfo {author} {\bibfnamefont {J.}~\bibnamefont
  {Rodr{\'\i}guez-Carvajal}},\ }\href@noop {} {\bibfield  {journal} {\bibinfo
  {journal} {Physica B}\ }\textbf {\bibinfo {volume} {192}},\ \bibinfo {pages}
  {55} (\bibinfo {year} {1993})}\BibitemShut {NoStop}%
\bibitem [{\citenamefont {Troyanchuk}\ and\ \citenamefont
  {Chobot}(1997)}]{troyanchuk97}%
  \BibitemOpen
  \bibfield  {author} {\bibinfo {author} {\bibfnamefont {I.~O.}\ \bibnamefont
  {Troyanchuk}}\ and\ \bibinfo {author} {\bibfnamefont {A.~N.}\ \bibnamefont
  {Chobot}},\ }\href@noop {} {\bibfield  {journal} {\bibinfo  {journal} {Cryst.
  Rep.}\ }\textbf {\bibinfo {volume} {42}},\ \bibinfo {pages} {983} (\bibinfo
  {year} {1997})}\BibitemShut {NoStop}%
\bibitem [{\citenamefont {Bochu}\ \emph {et~al.}(1980)\citenamefont {Bochu},
  \citenamefont {Buevoz}, \citenamefont {Chenavas}, \citenamefont {Collomb},
  \citenamefont {Joubert},\ and\ \citenamefont {Marezio}}]{bochu80}%
  \BibitemOpen
  \bibfield  {author} {\bibinfo {author} {\bibfnamefont {B.}~\bibnamefont
  {Bochu}}, \bibinfo {author} {\bibfnamefont {J.~L.}\ \bibnamefont {Buevoz}},
  \bibinfo {author} {\bibfnamefont {J.}~\bibnamefont {Chenavas}}, \bibinfo
  {author} {\bibfnamefont {A.}~\bibnamefont {Collomb}}, \bibinfo {author}
  {\bibfnamefont {J.~C.}\ \bibnamefont {Joubert}}, \ and\ \bibinfo {author}
  {\bibfnamefont {M.}~\bibnamefont {Marezio}},\ }\href@noop {} {\bibfield
  {journal} {\bibinfo  {journal} {Solid State Comm.}\ }\textbf {\bibinfo
  {volume} {36}},\ \bibinfo {pages} {133} (\bibinfo {year} {1980})}\BibitemShut
  {NoStop}%
\bibitem [{\citenamefont {Przenioslo}\ \emph {et~al.}(2002)\citenamefont
  {Przenioslo}, \citenamefont {Sosnowska}, \citenamefont {Suard}, \citenamefont
  {Hewat},\ and\ \citenamefont {Fitch}}]{przenioslo02}%
  \BibitemOpen
  \bibfield  {author} {\bibinfo {author} {\bibfnamefont {R.}~\bibnamefont
  {Przenioslo}}, \bibinfo {author} {\bibfnamefont {I.}~\bibnamefont
  {Sosnowska}}, \bibinfo {author} {\bibfnamefont {E.}~\bibnamefont {Suard}},
  \bibinfo {author} {\bibfnamefont {A.}~\bibnamefont {Hewat}}, \ and\ \bibinfo
  {author} {\bibfnamefont {A.~N.}\ \bibnamefont {Fitch}},\ }\href@noop {}
  {\bibfield  {journal} {\bibinfo  {journal} {J.Phys.: Condens. Matter}\
  }\textbf {\bibinfo {volume} {14}},\ \bibinfo {pages} {5747} (\bibinfo {year}
  {2002})}\BibitemShut {NoStop}%
\bibitem [{\citenamefont {Przenioslo}\ \emph {et~al.}(2004)\citenamefont
  {Przenioslo}, \citenamefont {Sosnowska}, \citenamefont {Suard}, \citenamefont
  {Hewat},\ and\ \citenamefont {Fitch}}]{przenioslo04}%
  \BibitemOpen
  \bibfield  {author} {\bibinfo {author} {\bibfnamefont {R.}~\bibnamefont
  {Przenioslo}}, \bibinfo {author} {\bibfnamefont {I.}~\bibnamefont
  {Sosnowska}}, \bibinfo {author} {\bibfnamefont {E.}~\bibnamefont {Suard}},
  \bibinfo {author} {\bibfnamefont {A.}~\bibnamefont {Hewat}}, \ and\ \bibinfo
  {author} {\bibfnamefont {A.~N.}\ \bibnamefont {Fitch}},\ }\href@noop {}
  {\bibfield  {journal} {\bibinfo  {journal} {Physica B}\ }\textbf {\bibinfo
  {volume} {344}},\ \bibinfo {pages} {358} (\bibinfo {year}
  {2004})}\BibitemShut {NoStop}%
\bibitem [{\citenamefont {Przenioslo}\ \emph {et~al.}(1999)\citenamefont
  {Przenioslo}, \citenamefont {Sosnowska}, \citenamefont {Hohlwein},
  \citenamefont {Hau\ss},\ and\ \citenamefont {Troyanchuk}}]{przenioslo99}%
  \BibitemOpen
  \bibfield  {author} {\bibinfo {author} {\bibfnamefont {R.}~\bibnamefont
  {Przenioslo}}, \bibinfo {author} {\bibfnamefont {I.}~\bibnamefont
  {Sosnowska}}, \bibinfo {author} {\bibfnamefont {D.}~\bibnamefont {Hohlwein}},
  \bibinfo {author} {\bibfnamefont {T.}~\bibnamefont {Hau\ss}}, \ and\ \bibinfo
  {author} {\bibfnamefont {I.~O.}\ \bibnamefont {Troyanchuk}},\ }\href@noop {}
  {\bibfield  {journal} {\bibinfo  {journal} {Solid State Comm.}\ }\textbf
  {\bibinfo {volume} {111}},\ \bibinfo {pages} {687} (\bibinfo {year}
  {1999})}\BibitemShut {NoStop}%
\bibitem [{\citenamefont
  {S$\mathrm{\acute{a}}$nchez-And$\mathrm{\acute{u}}$jar}\ \emph
  {et~al.}(2009)\citenamefont
  {S$\mathrm{\acute{a}}$nchez-And$\mathrm{\acute{u}}$jar}, \citenamefont
  {Y$\mathrm{\acute{a}\tilde{n}}$ez-Vilar}, \citenamefont {Biskup},
  \citenamefont {Castro-Garc{\'\i}a}, \citenamefont {Mira}, \citenamefont
  {Rivas},\ and\ \citenamefont
  {Se$\mathrm{\tilde{n}}$ar{\'\i}s-Rodr{\'\i}guez}}]{sanchezandujar09}%
  \BibitemOpen
  \bibfield  {author} {\bibinfo {author} {\bibfnamefont {M.}~\bibnamefont
  {S$\mathrm{\acute{a}}$nchez-And$\mathrm{\acute{u}}$jar}}, \bibinfo {author}
  {\bibfnamefont {S.}~\bibnamefont {Y$\mathrm{\acute{a}\tilde{n}}$ez-Vilar}},
  \bibinfo {author} {\bibfnamefont {N.}~\bibnamefont {Biskup}}, \bibinfo
  {author} {\bibfnamefont {S.}~\bibnamefont {Castro-Garc{\'\i}a}}, \bibinfo
  {author} {\bibfnamefont {J.}~\bibnamefont {Mira}}, \bibinfo {author}
  {\bibfnamefont {J.}~\bibnamefont {Rivas}}, \ and\ \bibinfo {author}
  {\bibfnamefont {M.~A.}\ \bibnamefont
  {Se$\mathrm{\tilde{n}}$ar{\'\i}s-Rodr{\'\i}guez}},\ }\href@noop {} {\bibfield
   {journal} {\bibinfo  {journal} {J. Mag. Mag. Mat.}\ }\textbf {\bibinfo
  {volume} {321}},\ \bibinfo {pages} {1739} (\bibinfo {year}
  {2009})}\BibitemShut {NoStop}%
\bibitem [{\citenamefont {Harris}\ \emph {et~al.}(1992)\citenamefont {Harris},
  \citenamefont {Kallin},\ and\ \citenamefont {Berlinsky}}]{harris92}%
  \BibitemOpen
  \bibfield  {author} {\bibinfo {author} {\bibfnamefont {A.~B.}\ \bibnamefont
  {Harris}}, \bibinfo {author} {\bibfnamefont {C.}~\bibnamefont {Kallin}}, \
  and\ \bibinfo {author} {\bibfnamefont {A.~J.}\ \bibnamefont {Berlinsky}},\
  }\href@noop {} {\bibfield  {journal} {\bibinfo  {journal} {Phys. Rev. B}\
  }\textbf {\bibinfo {volume} {45}},\ \bibinfo {pages} {2899} (\bibinfo {year}
  {1992})}\BibitemShut {NoStop}%
\bibitem [{\citenamefont {Sergienko}\ and\ \citenamefont
  {Dagotto}(2006)}]{sergienko06}%
  \BibitemOpen
  \bibfield  {author} {\bibinfo {author} {\bibfnamefont {I.~A.}\ \bibnamefont
  {Sergienko}}\ and\ \bibinfo {author} {\bibfnamefont {E.}~\bibnamefont
  {Dagotto}},\ }\href@noop {} {\bibfield  {journal} {\bibinfo  {journal} {Phys.
  Rev. B}\ }\textbf {\bibinfo {volume} {73}},\ \bibinfo {pages} {094434}
  (\bibinfo {year} {2006})}\BibitemShut {NoStop}%
\bibitem [{\citenamefont {Katsura}\ \emph {et~al.}(2005)\citenamefont
  {Katsura}, \citenamefont {Nagaosa},\ and\ \citenamefont
  {Balatsky}}]{katsura05}%
  \BibitemOpen
  \bibfield  {author} {\bibinfo {author} {\bibfnamefont {H.}~\bibnamefont
  {Katsura}}, \bibinfo {author} {\bibfnamefont {N.}~\bibnamefont {Nagaosa}}, \
  and\ \bibinfo {author} {\bibfnamefont {A.~V.}\ \bibnamefont {Balatsky}},\
  }\href@noop {} {\bibfield  {journal} {\bibinfo  {journal} {Phys. Rev. Lett.}\
  }\textbf {\bibinfo {volume} {95}},\ \bibinfo {pages} {057205} (\bibinfo
  {year} {2005})}\BibitemShut {NoStop}%
\bibitem [{\citenamefont {Kimura}\ \emph {et~al.}(2006)\citenamefont {Kimura},
  \citenamefont {Lashley},\ and\ \citenamefont {Ramirez}}]{kimura06}%
  \BibitemOpen
  \bibfield  {author} {\bibinfo {author} {\bibfnamefont {T.}~\bibnamefont
  {Kimura}}, \bibinfo {author} {\bibfnamefont {J.~C.}\ \bibnamefont {Lashley}},
  \ and\ \bibinfo {author} {\bibfnamefont {A.~P.}\ \bibnamefont {Ramirez}},\
  }\href@noop {} {\bibfield  {journal} {\bibinfo  {journal} {Phys. Rev. B}\
  }\textbf {\bibinfo {volume} {73}},\ \bibinfo {pages} {220401(R)} (\bibinfo
  {year} {2006})}\BibitemShut {NoStop}%
\bibitem [{\citenamefont {Arima}(2007)}]{arima07}%
  \BibitemOpen
  \bibfield  {author} {\bibinfo {author} {\bibfnamefont {T.}~\bibnamefont
  {Arima}},\ }\href@noop {} {\bibfield  {journal} {\bibinfo  {journal} {J.
  Phys. Soc. Jpn.}\ }\textbf {\bibinfo {volume} {76}},\ \bibinfo {pages}
  {073702} (\bibinfo {year} {2007})}\BibitemShut {NoStop}%
\bibitem [{\citenamefont {Kenzelmann}\ \emph {et~al.}(2007)\citenamefont
  {Kenzelmann}, \citenamefont {Lawes}, \citenamefont {Harris}, \citenamefont
  {Gasparovic}, \citenamefont {Broholm}, \citenamefont {Ramirez}, \citenamefont
  {Jorge}, \citenamefont {Jaime}, \citenamefont {Park}, \citenamefont {Huang},
  \citenamefont {Shapiro},\ and\ \citenamefont {Demianets}}]{kenzelmann07}%
  \BibitemOpen
  \bibfield  {author} {\bibinfo {author} {\bibfnamefont {M.}~\bibnamefont
  {Kenzelmann}}, \bibinfo {author} {\bibfnamefont {G.}~\bibnamefont {Lawes}},
  \bibinfo {author} {\bibfnamefont {A.~B.}\ \bibnamefont {Harris}}, \bibinfo
  {author} {\bibfnamefont {G.}~\bibnamefont {Gasparovic}}, \bibinfo {author}
  {\bibfnamefont {C.}~\bibnamefont {Broholm}}, \bibinfo {author} {\bibfnamefont
  {A.~P.}\ \bibnamefont {Ramirez}}, \bibinfo {author} {\bibfnamefont {G.~A.}\
  \bibnamefont {Jorge}}, \bibinfo {author} {\bibfnamefont {M.}~\bibnamefont
  {Jaime}}, \bibinfo {author} {\bibfnamefont {S.}~\bibnamefont {Park}},
  \bibinfo {author} {\bibfnamefont {Q.}~\bibnamefont {Huang}}, \bibinfo
  {author} {\bibfnamefont {A.~Y.}\ \bibnamefont {Shapiro}}, \ and\ \bibinfo
  {author} {\bibfnamefont {L.~A.}\ \bibnamefont {Demianets}},\ }\href@noop {}
  {\bibfield  {journal} {\bibinfo  {journal} {Phys. Rev. Lett.}\ }\textbf
  {\bibinfo {volume} {98}},\ \bibinfo {pages} {267205} (\bibinfo {year}
  {2007})}\BibitemShut {NoStop}%
\end{thebibliography}%

\end{document}